# The generalized e-bundle

Leo Egghe[1] and Ronald Rousseau[2,3] [*]

[1] Hasselt University, 3500 Hasselt, Belgium

leo.egghe@uhasselt.be

ORCID: 0000-0001-8419-2932

[2] KU Leuven, MSI, Facultair Onderzoekscentrum ECOOM,

Naamsestraat 61, 3000 Leuven, Belgium

ronald.rousseau@kuleuven.be   &

[3] University of Antwerp, Faculty of Social Sciences,

Middelheimlaan 1, 2020 Antwerp, Belgium

ronald.rousseau@uantwerpen.be

ORCID: 0000-0002-3252-2538

**Abstract**

In previous work, we introduced the notion of an impact bundle, showing how e.g., the h-index and the g-index can lead to such a bundle. Here we extend the set of impact bundles by a new impact bundle, based on Zhang's e-index. It is, moreover, shown that some other plausible definitions do not lead to an impact bundle.

## 1. Introduction

The field of scientometrics/informetrics studies aspects of the science of science. It is not just a theoretical field, but also has important applications in research evaluation, as an aid to decision-makers. Yet, this contribution is a purely theoretical (mathematical) study. It continues the earlier work of the authors (Egghe, 2022a, 2022b; Egghe & Rousseau, 2022, 2023).

In this article, we study a generalization of Zhang's e-index in a bundle context. Hence, we first recall the definition of a bundle.

Definition. Bundles (Egghe & Rousseau, 2022a).

Let $\mathbf{Z} \subset \mathbf{U} = \{Z; Z: [0,T] \to \mathbf{R}^+, \text{ continuous and decreasing}\}$. Then **m**: $\mathbf{Z} \times [0,+\infty] \to \mathbf{R}^+$: $(Z,\theta) \to \mathbf{m}_Z(\theta)$ is a bundle if moreover there exists for every $Z$ in $\mathbf{Z}$, a continuous injection, $\psi_Z: [0,T] \to [0,+\infty]$: $x \to \psi_Z(x) = \theta$. Hence, we may refer to a bundle **B** as the couple (**m**, $\psi$). Given Z we refer to all θ-values for which **m**(Z,.) is defined as admissible θ values. This set is denoted as $Q_Z$. If we study two functions Z and Y at the same time, e.g., when studying the relation Y ≥ Z, admissible θ values are values in the intersection $Q_Y \cap Q_Z$.

As a continuous injective real function on an interval is strictly monotonous (Hairer & Wanner, 2008), we have that $\psi_Z$ is either strictly increasing or strictly decreasing. If we study two functions Z and Y at the same time, we always assume that either $\psi_Z$ and $\psi_Y$ are both increasing or are both decreasing.

Definition. An impact bundle (Egghe & Rousseau, 2022a)

A bundle **B** = (**m**, $\psi$) is called an impact bundle if it satisfies the following four requirements.

(AX.1). For all admissible θ, **m**(**0,** θ) = 0, where **0** is the zero function.

This axiom is trivially satisfied if **0** $\notin$ **Z**.

(AX.2). For all Y, Z ∈ **Z,** and all admissible θ : Y ≥ Z ⇒ $\mathbf{m}_Y(\theta) \geq \mathbf{m}_Z(\theta)$.

(AX.3) A bundle (**m**, $\psi$) on **Z** meets the requirements of axiom (AX.3) iff for all a in ]0, T[, and for all Y, Z in **Z**: Y > Z on [0, a] ⇒ $m_Y(\theta) > m_Z(\theta)$ for all admissible θ in $\psi_Y([0,a]) \cup \psi_Z([0,a])$.

(AX.4) For all a ∈ ]0,T[, and for all Y,Z ∈ **Z**, such that Y=Z on [0,a] we have:

$$\psi_Z|_{[0,a]} = \psi_Y|_{[0,a]}$$

and **m**$_Y$(θ) = **m**$_Z$(θ), for $\theta \in \psi_Z([0,a]) = \psi_Y([0,a])$

Three examples of bundles

A. The average number of items in the first sources

In this case we take any set **Z** ⊂ **U** , Q$_Z$ = [0,T] for all Z,  ψ$_Z$: [0,T] → [0,T]: x→ θ = ψ$_Z$(x) = x (strictly increasing and the same for each Z) and μ: **Z** ⊂ **U** × [0,T] → **R**$^+$: (Z,θ) → μ$_Z$(θ) = $\frac{1}{\theta}\int_0^\theta Z(s)ds$; if θ = 0 then μ$_Z$(0)=Z(0).

B. Total number of items in an interval starting in 0

This is similar to the previous example. We again use any set **Z** ⊂ **U** , Q$_Z$ = [0,T] for all Z, and use the same function ψ$_Z$: [0,T] → [0,T]: x→ θ = ψ$_Z$(x) = x. Then we define I: **Z** ⊂ **U** × [0,T] → **R**$^+$: (Z,θ) → I$_Z$(θ) = $\int_0^\theta Z(s)ds$.

These examples are rather elementary. Next, we consider a more complicated one.

C. The generalized h-index

We take **Z** = $\{Z; Z: [0,T] \to$ **R**$^+,$ continuous, decreasing and Z(x) > 0 for $x \in [0,T[\}$, ψ$_Z$: [0,T] → Q$_Z$ = [Z(T)/T, +∞]: x→ θ = ψ$_Z$(x) = Z(x)/x (strictly decreasing and depending on Z) and m: **Z** ⊂ **U** × [0,+∞] → **R**$^+$: (Z,θ) → m$_Z$θ = h$_θ$(Z), where Z(h$_θ$(Z)) = θ h$_θ$(Z).

More examples of bundles can be found in (Egghe & Rousseau, 2022a), where it is moreover shown that the three bundles mentioned above, and many more are impact bundles.

## 2. The e-index

Zhang (2009) introduced the e-index, discrete and continuous, as

$e = \sqrt{R^2 - h^2}$

where the symbol e refers to excess citations; h is de classical h-index (Hirsch, 2005) and R is the R-index as introduced in (Jin et al., 2007). An advantage of the e-index, with respect to the h-index, is the fact that it takes all citations of the most-cited sources into account. Further studies on the discrete e-index can be found e.g., in (Zhang, 2010; Yuan et al., 2014).

In this article we will study in detail a generalized continuous e-index, somewhat similar to the generalized h- and g-indices (van Eck & Waltman, 2008; Egghe & Rousseau, 2019).

## 3. The continuous generalized e-index

From now on we work in the set $U_s =$

$\{Z; Z: [0,T] \to \mathbf{R}^+,$ continuous and strictly decreasing$\}$. As each Z in $U_s$ is strictly decreasing, $U_s$ does not contain constant functions.

3.1 Definition.

We define the **e**-bundle as (**e**, ψ) with **e**: $U_s \times [0,+\infty] \to \mathbf{R}^+$: (Z,θ) → $e_\theta(Z) = \int_0^{Z^{-1}(\theta)} (Z(s) - \theta)\, ds$ ,and   ψ$_Z$: [0,T] → [Z(T), Z(0)] ⊂ [0,+∞]: x→ ψ$_Z$(x) = θ = Z(x).

For each admissible value of θ we have a generalized e-index and if θ = h(Z), we obtain Zhang's e-index of the function Z.

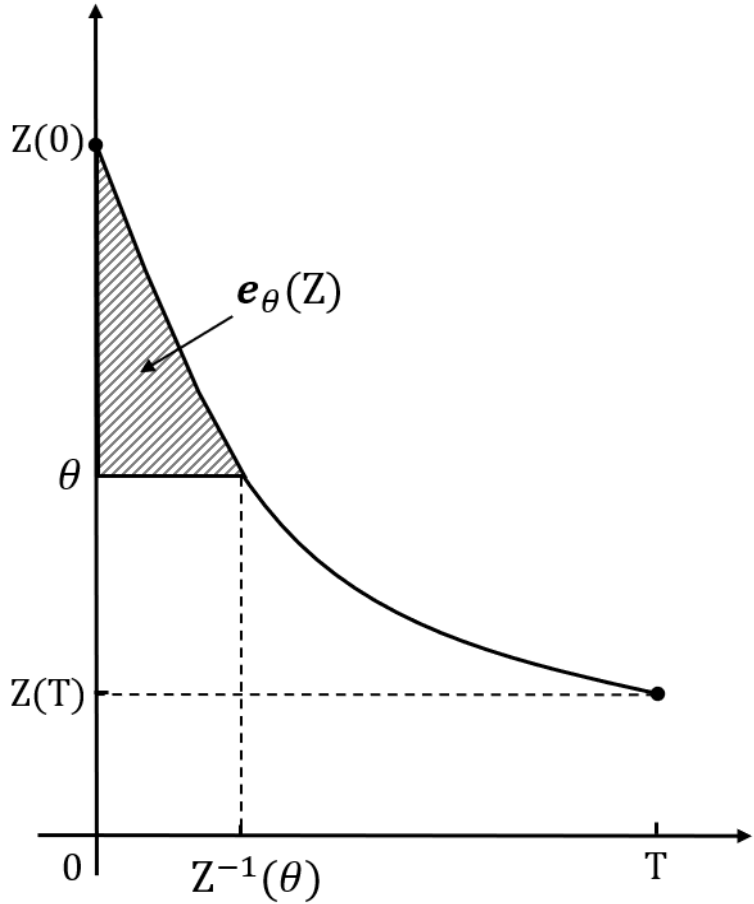

Fig.1 An illustration of $e_\theta(Z)$ for a general function Z in $U_s$

3.2 Theorem 1

The **e**-bundle is an impact bundle

Proof.

(AX.1) is satisfied because $0 \notin U_s$.

(AX.2). Assume that Y, Z ∈ $U_s$ and that Y ≥ Z on [0,T]. We have to show that then $e_\theta(Y) \geq e_\theta(Z)$ for all admissible values of θ. If Y ≥ Z then Y(Z⁻¹(θ)) ≥ Z(Z⁻¹(θ))=θ. As Y is strictly decreasing, we know that, x < Z⁻¹(θ) implies that Y(x) > Y(Z⁻¹(θ)) ≥ θ. Now we know that Y(Y⁻¹(θ)) = θ hence it is not true that Y⁻¹(θ) < Z⁻¹(θ) and thus Z⁻¹(θ) ≤ Y⁻¹(θ).

Consequently, for all admissible θ, $e_\theta(Y) = \int_0^{Y^{-1}(\theta)}(Y(s) - \theta)\,ds \geq \int_0^{Z^{-1}(\theta)}(Z(s) - \theta)\,ds = e_\theta(Z)$.

(AX.3). Given a ∈ ]0,T[ and Y, Z ∈ **$U_s$** such that Y > Z on [0,a]. Then we have to show that for all $\theta \in \psi_Y([0,a]) \cup \psi_Z([0,a])$, **e**₀(Y) > **e**₀(Z). We first note that $\psi_Y([0,a]) \cup \psi_Z([0,a])$ = [Z(a), Z(0)] ∪ [Y(a), Y(0)] = [Z(a),Y(0)], because Y > Z on [0,a]. We consider now two cases:

(I): θ ∈ [Z(a), Z(0)]. Then there exists $x_1$ ∈ [0,a] such that Z($x_1$) = θ < Y($x_1$) and as $Y^{-1}$ is strictly decreasing we have $Z^{-1}$(θ) = $x_1$ < $Y^{-1}$(θ).

(II): θ ∈ [Y(a), Y(0)]. Then there exists $x_2$ ∈ [0,a] such that Y($x_2$) = θ > Z($x_2$) and as $Z^{-1}$ is strictly decreasing we have $Y^{-1}$(θ) = $x_2$ > $Z^{-1}$(θ).

From (I) and (II) we conclude that for all $\theta \in \psi_Y([0,a]) \cup \psi_Z([0,a])$: $Y^{-1}$(θ) > $Z^{-1}$(θ).

In case (I) we have

$$\mathbf{e}_\theta(Y) = \int_0^{Y^{-1}(\theta)}(Y(s)-\theta)\,ds > \int_0^{Z^{-1}(\theta)}(Y(s)-\theta)\,ds > \mathbf{e}_\theta(Z) = \int_0^{Z^{-1}(\theta)}(Z(s)-\theta)\,ds$$

, because θ ∈ [Z(a), Z(0)] and hence $Z^{-1}$(θ) ∈ [0,a], where Y > Z.

In case (II) we have

$$\mathbf{e}_\theta(Z) = \int_0^{Z^{-1}(\theta)}(Z(s)-\theta)\,ds < \int_0^{Y^{-1}(\theta)}(Z(s)-\theta)\,ds < \mathbf{e}_\theta(Y) = \int_0^{Y^{-1}(\theta)}(Y(s)-\theta)\,ds$$

, because θ ∈ [Y(a), Y(0)] and hence $Y^{-1}$(θ) ∈ [0,a], where Y > Z. This proves (AX.3) for $\theta \in \psi_Y([0,a]) \cup \psi_Z([0,a])$.

(AX.4). Consider a ∈ ]0,T[, and Y,Z ∈ **$U_s$**, such that Y=Z on [0,a] then we have by definition that

$$\psi_Z|_{[0,a]} = \psi_Y|_{[0,a]}$$

and **e**_Y(θ) = **e**_Z(θ), for $\theta \in \psi_Z([0,a]) = \psi_Y([0,a])$

This ends the proof that the **e**-bundle is an impact bundle □

## 3. Definitions

In earlier articles, we introduced several measures of impact. We recall these here. As there exist many measures in the literature which are referred to as impact measures, we call the first one "impact measure in the sense of Egghe".

Impact measures in the sense of Egghe (Egghe, 2022a)

A function m defined on a set **Z** ⊂ **U** = {Z ‖ Z: [0, T] → **R**⁺, continuous and decreasing}, m: **Z** ⊂ **U** → **R**⁺: Z → m(Z), is an impact measure in the sense of Egghe if it meets the following three requirements:

(I) m(Z) = 0 if and only if Z = 0.

(II) Y ≥ X ⇒ m(Y) ≥ m(X) and Y = X ⇒ m(Y) = m(X).

(III) ∀ X ∈ **Z**, ∃ $a_X$ ∈ ]0,T[ such that: for all Y, Z in **Z**, (Y >> Z on [0, min($a_Y$,$a_Z$)] implies that m(Y) > m(Z)).

Here Y >> Z on an interval [0,b] means that for all x ∈ [0,b], Y(x) > Z(x).

From this definition and the previous theorem, we immediately have the following corollary.

Corollary

For all admissible θ, $e_\theta$ is an impact measure in the sense of Egghe.

Proof. This follows from the corollary of Theorem 1 in (Egghe & Rousseau, 2022a).

Strong impact measures (Egghe & Rousseau, 2022a)

We define for every Z ∈ **U**, the average function $\mu_Z$ as:

$$\mu_Z: [0,T] \to \mathbf{R}^+: x \to \frac{1}{x}\int_0^x Z(s)ds \text{ and } \mu_Z(0) = Z(0)$$

The graph of $\mu_Z$ is referred to as the strong impact curve of Z.

Definition. A strong impact measure on **Z** ⊂ **U** is a function m from **Z** ⊂ **U** to **R**⁺, satisfying the following four axioms.

(ax.1). m(Z) = 0 if and only if Z = **0** (the null function, mapping each point to the value 0; it may or may not belong to **Z**).

(ax.2). For all Y, Z ∈ **Z** ⊂ **U**: Z ≤ Y ⇒ m(Z) ≤ m(Y).

(ax.3). For all Y, Z ∈ **Z** ⊂ **U**: $\mu_Z < \mu_Y$ on $[0,T[ \Rightarrow m(Z) < m(Y)$

(ax.4). For every X in **Z** ⊂ **U**, there exists $a_X$ in ]0,T[ such that, for all Y, Z in **Z**: (Y=Z on [0,min($a_Y$, $a_Z$)], implies m(Y) = m(Z)).

Global impact measures (Egghe & Rousseau, 2023)

Let $\boldsymbol{U_0} = \{Z \in U; Z > 0 \text{ on } [0, T[\}$. Then a function m: $\boldsymbol{U_0} \to \mathbb{R}^+$ such that $\forall Z, Y \in \boldsymbol{U_0}, Z \neq Y$: $(Z \prec Y \Rightarrow m(Z) < m(Y))$ is called a global impact measure. Here $Z \prec Y$ iff $\forall x \in [0,T]: I_Z(x) = \int_0^x Z(s)ds \leq I_Y(x) = \int_0^x Y(s)ds$.

Theorem 2. For all admissible $\theta \neq Z(T)$, $\boldsymbol{e_\theta}$ is a strong impact measure.

Proof. We have to check four axioms. The first two are trivial. Now we check axiom three, (ax.3). Given $\mu_Z < \mu_Y$ on [0,T[ we have to show that for all admissible $\theta \neq Z(T)$, $\boldsymbol{e_\theta}(Z) < \boldsymbol{e_\theta}(Y)$. If $Z^{-1}(\theta) \leq Y^{-1}(\theta)$ then this is easy to show as

$\boldsymbol{e_\theta}(Z) = \int_0^{Z^{-1}(\theta)}(Z(s) - \theta)\, ds < \int_0^{Z^{-1}(\theta)}(Y(s) - \theta)\, ds$ (this follows from $\mu_Z < \mu_Y$ on [0,T[ and $Z^{-1}(\theta) \neq T$] $< \int_0^{Y^{-1}(\theta)}(Y(s) - \theta)\, ds$ (this follows from the fact that $Y > \theta$ on $[0, Y^{-1}(\theta)]$ and that Y is strictly decreasing) $= \boldsymbol{e_\theta}(Y)$.

Now we consider the case that $Z^{-1}(\theta) > Y^{-1}(\theta)$. Then

$$\boldsymbol{e_\theta}(Y) - \boldsymbol{e_\theta}(Z) = \int_0^{Y^{-1}(\theta)}(Y(s) - \theta)\, ds - \int_0^{Z^{-1}(\theta)}(Z(s) - \theta)\, ds$$

$= \int_0^{Y^{-1}(\theta)}(Y(s))\, ds - \theta.Y^{-1}(\theta) - \int_0^{Z^{-1}(\theta)}(Z(s))\, ds + \theta.Z^{-1}(\theta)$

$= \int_0^{Y^{-1}(\theta)}(Y(s))\, ds - \theta(Y^{-1}(\theta) - Z^{-1}(\theta)) - \int_0^{Y^{-1}(\theta)}(Z(s))\, ds - \int_{Y^{-1}(\theta)}^{Z^{-1}(\theta)}(Z(s))\, ds$

$= \int_0^{Y^{-1}(\theta)}(Y(s) - Z(s))\, ds - \left[\int_{Y^{-1}(\theta)}^{Z^{-1}(\theta)}(Z(s))\, ds - \theta(Y^{-1}(\theta) - Z^{-1}(\theta))\right]$

The second term is equal to:

$$\int_{Y^{-1}(\theta)}^{Z^{-1}(\theta)}(Z(s) - \theta)\, ds$$

$< \int_{Y^{-1}(\theta)}^{Z^{-1}(\theta)}(Z(s) - Y(s))\, ds$ as $Y < \theta$ on $[Y^{-1}(\theta), Z^{-1}(\theta)]$.

Hence:

$$\boldsymbol{e_\theta}(Y) - \boldsymbol{e_\theta}(Z) > \int_0^{Y^{-1}(\theta)}(Y(s) - Z(s))\, ds - \int_{Y^{-1}(\theta)}^{Z^{-1}(\theta)}(Z(s) - Y(s))\, ds$$

$$= \int_0^{Y^{-1}(\theta)}(Y(s) - Z(s))\,ds + \int_{Y^{-1}(\theta)}^{Z^{-1}(\theta)}(Y(s) - Z(s))\,ds = \int_0^{Z^{-1}(\theta)}(Y(s) - Z(s))\,ds > 0$$

(as $\mu_Z < \mu_Y$ on [0,T[ and $Z^{-1}(\theta) \neq T$. This proves (AX.3)

Finally, we have to show (ax.4): For every X in $\mathbf{Z} \subset \mathbf{U}$, there exists $a_X$ in ]0,T[ such that, for all Y, Z in $\mathbf{Z}$: (Y=Z on [0,min($a_Y$, $a_Z$)], implies $e_\theta(Y) = e_\theta(Z)$ ). This axiom holds because we can take $a_Z = Z^{-1}(\theta) = a_Y = Y^{-1}(\theta)$ and hence, trivially $e_\theta(Y) = e_\theta(Z)$.

Corollary.

For all admissible $\theta \neq Z(T)$, $e_\theta$ is an impact measure in the sense of Egghe.

This follows immediately from Proposition 2 in (Egghe & Rousseau, 2022a).

Proposition

The measure $e_\theta$ is not a global impact measure for all admissible $\theta \neq Z(T)$.

Proof. Consider the functions Y and Z as shown in Fig 2. These functions are strictly decreasing and continuous.

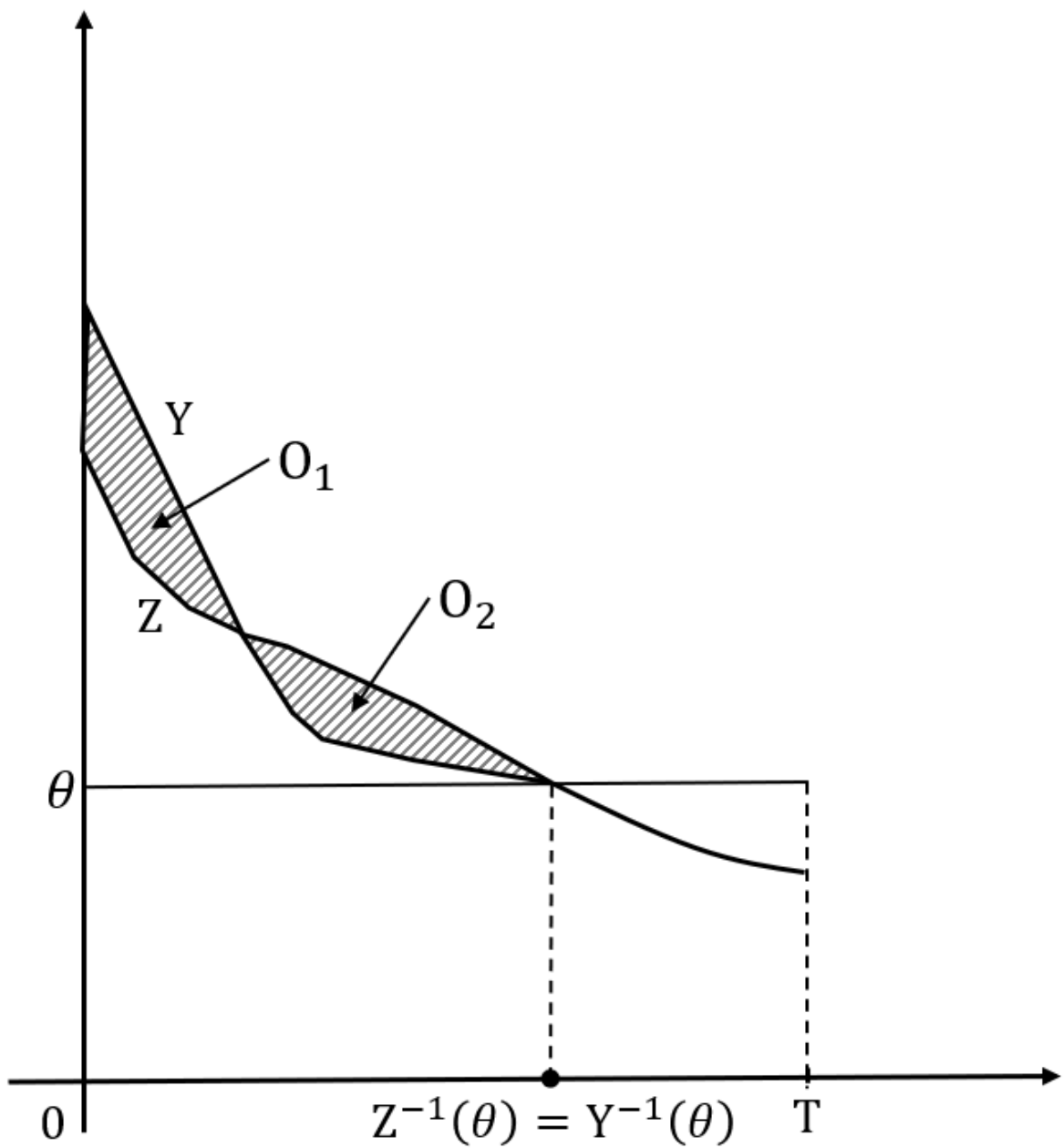

Fig. 2. Functions Y and Z used to show that $e_\theta$ is not a global impact measure for all admissible $\theta \neq Z(T)$

In Fig.2 we let θ be the ordinate of the second intersection point of Y and Z. Hence $Y^{-1}(\theta) = Z^{-1}(\theta)$. The functions Y and Z are different, but $O_1 = O_2$. For completeness' sake we set Y = Z on the interval $[Y^{-1}(\theta) = Z^{-1}(\theta), T]$. Hence, Z -<≠ Y. Then we have:

$$e_\theta(Z) = \int_0^{Z^{-1}(\theta)} (Z(s) - \theta)\, ds = \int_0^{Z^{-1}(\theta)} Z(s)ds - \theta.Z^{-1}(\theta)$$

$$= \int_0^{Y^{-1}(\theta)} Y(s)ds - \theta.Y^{-1}(\theta) = e_\theta(Y)$$

This shows that $e_\theta$ is not a global impact measure.

## 4. Convergence of the e-bundle

In (Egghe, 2022b) we studied the convergence of impact measures and impact bundles. This led to a classification of impact bundles. From this investigation we recall the following definitions:

Definition. Pointwise convergence

We say that $(Z_n)_n \to Z$, pointwise, with all $Z_n$ in **U** iff

$$\forall x \in [0,T]: \lim_{n \to \infty} Z_n(x) = Z(x) \text{ in } \mathbf{R}^+.$$

Here we can make a distinction between the case that $Z \in \mathbf{U}$ (pointwise convergence in **U**) and the case that $Z \notin \mathbf{U}$. In the latter case we will say that there is pointwise convergence on **U**.

Definition: uniform convergence

We say that $(Z_n)_n \to Z$, uniformly in **U** iff

$\forall \varepsilon > 0,\ \exists n_0$ such that, $\forall x \in [0,T]$, $n \geq n_0 \Rightarrow |Z_n(x) - Z(x)| < \varepsilon$, with $Z \in \mathbf{U}$

The point is that $n_0$ does not depend on x. We further note that the uniform limit of continuous functions is continuous (Apostol, 1967, 11.3). It is obvious that when $(Z_n)_n \to Z$, uniformly in **U** then also $(Z_n)_n \to Z$, pointwise in **U**.

We next mention three propositions. The first is well-known, the second one (Dini's second theorem) is not so well-known but was already used in a previous publication (Egghe, 2022b), while we could not find a reference for the third one and hence provide a proof here.

Proposition 1. If the function Z is strictly decreasing and continuous on [0,T] then $Z^{-1}$ is uniformly continuous on [Z(T), Z(0)].

A proof can be found in (De Lillo, 1982; p. 119).

Proposition 2. (Dini's second theorem) Let $(f_n)_n$ be a sequence of increasing or decreasing real functions, defined on the interval [a,b], such that $(f_n)_n$ tends pointwise to a continuous function f, then $(f_n)_n$ tends to f uniformly.

Proposition 3. Assume that for all n, $Z_n$ is continuous and strictly decreasing on [0,T]. Then

$(Z_n)_n$ tends to Z pointwise, with Z continuous and strictly decreasing, implies that $(Z_n)^{-1}$ converges uniformly to $Z^{-1}$.

Proof. By Proposition 1 we know that $Z^{-1}$ is uniformly continuous on [Z(T), Z(0)]. Hence, $\forall \varepsilon > 0, \exists \delta > 0$ such that $|x - y| \leq \delta$ (x,y ∈ [Z(T), Z(0)]), implies that $|Z^{-1}(x)-Z^{-1}(y)| < \varepsilon$. Consequently, we have:

$|Z^{-1}(x)-Z^{-1}(x+\delta)| < \varepsilon$ and $|Z^{-1}(x)-Z^{-1}(x-\delta)| < \varepsilon$  (*)

By the second Dini theorem we know that $Z^{-1}$ being continuous implies that $(Z_n^{-1})_n$ tends to $Z^{-1}$ uniformly. Hence, for δ > 0 whose existence was shown above, we know that there exists $n_0$ such that $n \geq n_0$ implies that for all y ∈ [Z(T), Z(0)]:

$$\left|Z_n(Z^{-1}(y)) - Z(Z^{-1}(y))\right| = \left|Z_n(Z^{-1}(y)) - y\right| < \delta$$

Hence,

$$\left|Z_n(Z^{-1}(x+\delta)) - (x+\delta)\right| < \delta \text{ and } \left|Z_n(Z^{-1}(x-\delta)) - (x-\delta)\right| < \delta.$$

From this we see that:

$$-\delta < Z_n(Z^{-1}(x+\delta)) - (x+\delta) < \delta$$

and

$$-\delta < Z_n(Z^{-1}(x-\delta)) - (x-\delta) < \delta$$

Applying $Z_n^{-1}$ and because all $Z_n$ are strictly decreasing, we obtain: for all n:

$$Z^{-1}(x+\delta) < Z_n^{-1}(x) \text{ and } Z^{-1}(x-\delta) > Z_n^{-1}(x)$$

Hence: $Z_n^{-1}(x) \in [Z^{-1}(x+\delta), Z^{-1}(x-\delta)]$

From (*) we know that $Z^{-1}(x) - \varepsilon < Z^{-1}(x+\delta)$ and $Z^{-1}(x) + \varepsilon > Z^{-1}(x-\delta)$. Consequently, $[Z^{-1}(x+\delta), Z^{-1}(x-\delta)] \subset [Z^{-1}(x) - \varepsilon, Z^{-1}(x) + \varepsilon]$. This implies that $Z_n^{-1}(x) \in [Z^{-1}(x) - \varepsilon, Z^{-1}(x) + \varepsilon]$ and thus $|Z_n^{-1}(x) - Z^{-1}(x)| < \varepsilon$, uniformly in x. □

Theorem 3Theorem 3

If $(Z_n)_n$ tends to Z pointwise in $U_s$ and if Z and all $Z_n$ are strictly decreasing and continuous, then $(e(Z_n))_n$ tends to $e(Z)$ uniformly.

Proof. For all admissible θ we have:

$$|e_\theta(Z_n) - e_\theta(Z)| = \left|\int_0^{Z_n^{-1}(\theta)}(Z_n(s)-\theta)\,ds - \int_0^{Z^{-1}(\theta)}(Z(s)-\theta)\,ds\right|$$

$$= \left|\int_0^{Z_n^{-1}(\theta)}(Z_n(s))ds - \theta Z_n^{-1}(\theta) - \int_0^{Z^{-1}(\theta)}(Z(s))ds + \theta Z^{-1}(\theta)\right|$$

$$= \left|\int_0^{Z^{-1}(\theta)}(Z_n(s))ds + \int_{Z^{-1}(\theta)}^{Z_n^{-1}(\theta)}(Z_n(s))ds - \theta Z_n^{-1}(\theta) - \int_0^{Z^{-1}(\theta)}Z(s)\,ds + \theta Z^{-1}(\theta)\right|$$

$$\leq \left|\int_0^{Z^{-1}(\theta)}(Z_n(s)-Z(s))ds\right| + \theta|Z_n^{-1}(\theta)-Z^{-1}(\theta)| + \left|\int_{Z^{-1}(\theta)}^{Z_n^{-1}(\theta)}(Z_n(s))ds\right|$$

As θ is admissible we know that θ ≤ Z(0). Moreover, $\left|\int_{Z^{-1}(\theta)}^{Z_n^{-1}(\theta)}(Z_n(s))ds\right| \leq M\,|Z_n^{-1}(\theta) - Z^{-1}(\theta)|$, where M = sup$_n$ ($Z_n(0)$) < +∞.

This shows, by proposition 3, that the second and third term in the above inequality converge uniformly to zero. Next, we focus on the first term. We know from (Egghe, 2022b, Theorem 8) that from $(Z_n)_n$ tending to Z pointwise in $U$ (hence also in $U_s$) it follows that $I_\theta(Z_n) = \int_0^\theta Z_n(s)ds$ tends to $I_\theta(Z) = \int_0^\theta Z(s)ds$ uniformly, hence also pointwise, in θ. Consequently, we also have that $\int_0^{Z^{-1}(\theta)} Z_n(s)ds$ tends to $\int_0^{Z^{-1}(\theta)} Z(s)ds$ pointwise. We further know that for all n, $\int_0^{Z^{-1}(\theta)} Z_n(s)ds$ decreases in θ, because $Z^{-1}$ is strictly decreasing in θ and $\int_0^{Z^{-1}(\theta)} Z(s)ds$ is continuous in θ (use proposition1).

Now we apply Dini's second theorem and obtain that also the first term converges uniformly in θ. This proves that $|e_\theta(Z_n) - e_\theta(Z)|$ tends to zero, uniformly in θ, and hence $(e(Z_n))_n$ tends to $e(Z)$ uniformly. □

Examples

Example 1.

Given T>0, then we form the set of functions

$$Z_S(x) = S\left(1 - \frac{x}{T}\right), x \in [0,T]$$

Then $Z_S^{-1}(\theta) = T\left(1 - \frac{\theta}{S}\right), \theta \in [0,S]$. The generalized e-index for this set of functions $Z_S$ is then, see Fig.3,

$$e_\theta(Z) = \frac{S-\theta}{2} \cdot Z_S^{-1}(\theta) = \frac{S-\theta}{2} \cdot T\left(1 - \frac{\theta}{S}\right) = \frac{T}{2S}(S-\theta)^2, \theta \in [0,S].$$

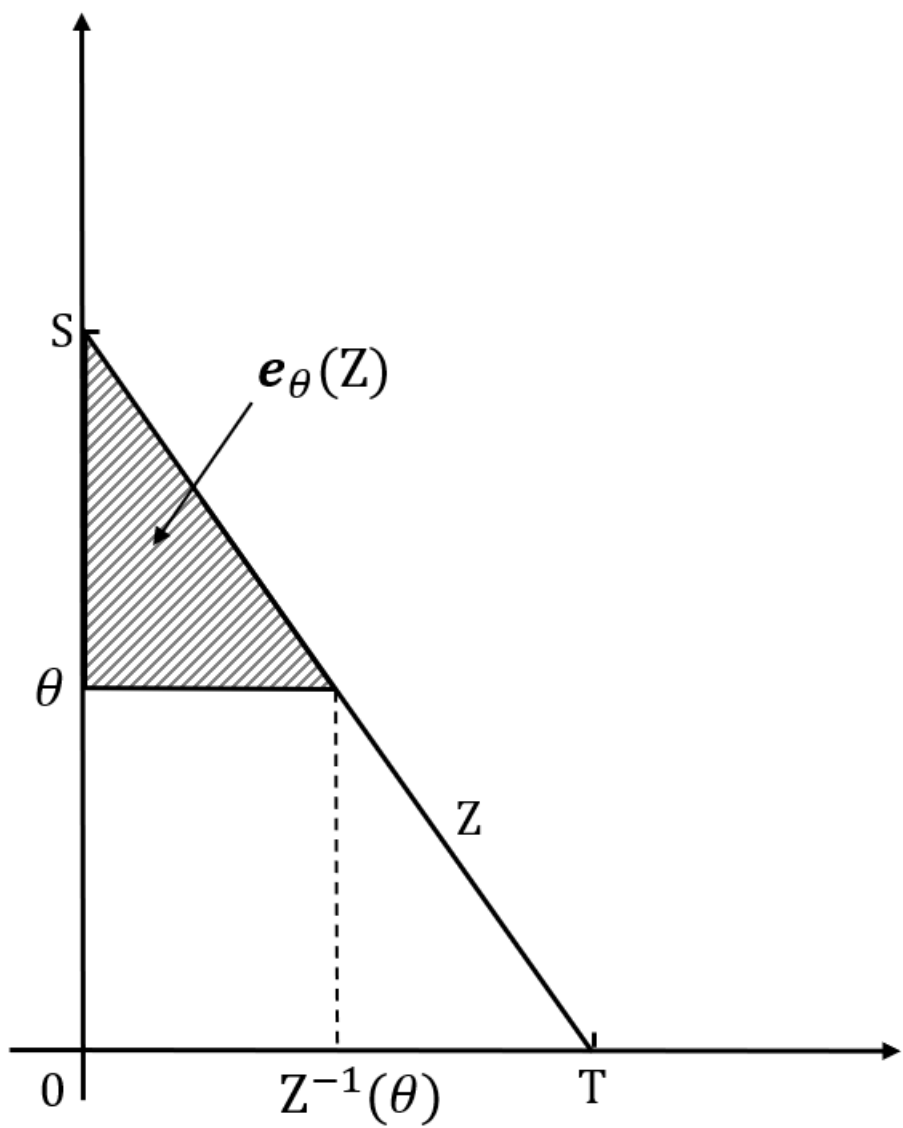

Fig. 3 Illustration of Example 1

Example 2.

Zipf's function (making sure integrals converge) on [0,T]

For 0 < β < 1, we take $Z_\beta(x) = \left(\frac{T}{x}\right)^\beta$, corresponding to the Lotka function with exponent α = (β+1)/β.

Then $Z_\beta^{-1}(\theta) = T\,\theta^{-1/\beta}$, $\theta \in [1, +\infty[$. Now the generalized e-index for this set of functions $Z_\beta$ is:

$$e_\theta(Z_\beta) = \int_0^{Z_\beta^{-1}(\theta)} (Z(s) - \theta)\,ds = \int_0^{T\,\theta^{-1/\beta}} \left(\left(\frac{T}{s}\right)^\beta - \theta\right) ds$$

$$= \left[T^\beta \frac{s^{-\beta+1}}{1-\beta} - \theta s\right]_0^{T\,\theta^{-1/\beta}} = T^\beta \frac{T^{-\beta+1}}{1-\beta} \theta^{(\beta-1)/\beta} - \theta\,T\,\theta^{-1/\beta}$$

$$= T\frac{\beta}{1-\beta}\,\theta^{1-1/\beta} = \frac{T}{\left(\frac{1}{\beta}-1\right)\theta^{\left(\frac{1}{\beta}\right)-1}}.$$

Rewriting this result using the corresponding Lotka α yields:

$$e_\theta(Z_\alpha) = \frac{T}{(\alpha-2)\theta^{\alpha-2}}$$

We note that this result is quite different from the generalized h, g or R index for Zipf functions.

Example 3

Consider $Z_n(x) = 1 - x^n$, $x \in [0,1]$. Then $Z_n$ is strictly decreasing and continuous for each n. The functions $Z_n$ tend pointwise to Z where $Z(x) = 1$, $x \in [0,1[$, and $Z(1) = 0$. As this function Z is not continuous, the theory exposed in our article does not apply.

## 5. Alternatives

While performing the research for this article we considered, possibly simpler, alternatives for the **e**-bundle. Here we provide two proposals, which, however, do not satisfy our requirements. Hence, we are convinced that within our framework the **e**-bundle is a relatively simple bundle generalization of Zhang's idea.

Alternative 1.

We define for each $Z \in \boldsymbol{U_s}$ and appropriate values of θ

$$n_\theta(Z) = \frac{1}{Z^{-1}(\theta)} \int_0^{Z^{-1}(\theta)} (Z(s) - \theta) ds$$

Yet, $n_\theta(Z)$ is not increasing in Z, which is a basic requirement for any impact measure. The following example shows this. Consider Fig. 4 with Y and Z strictly decreasing and continuous. The area between the curve of Y and the curve of Z, situated above [0, Y$^{-1}$(θ)] is called A. This area can be made as small as we want. We see that Y$^{-1}$(θ) >> Z$^{-1}$(θ) and Z < Y on [0,T[. Yet, $n_\theta(Y) < n_\theta(Z)$. Indeed,

$\int_0^{Y^{-1}(\theta)} (Y(s) - \theta) ds = \int_0^{Z^{-1}(\theta)} (Z(s) - \theta) ds$ + A. As Y$^{-1}$(θ) >> Z$^{-1}$(θ) we have that $n_\theta(Y) < n_\theta(Z)$ if A is taken small enough.

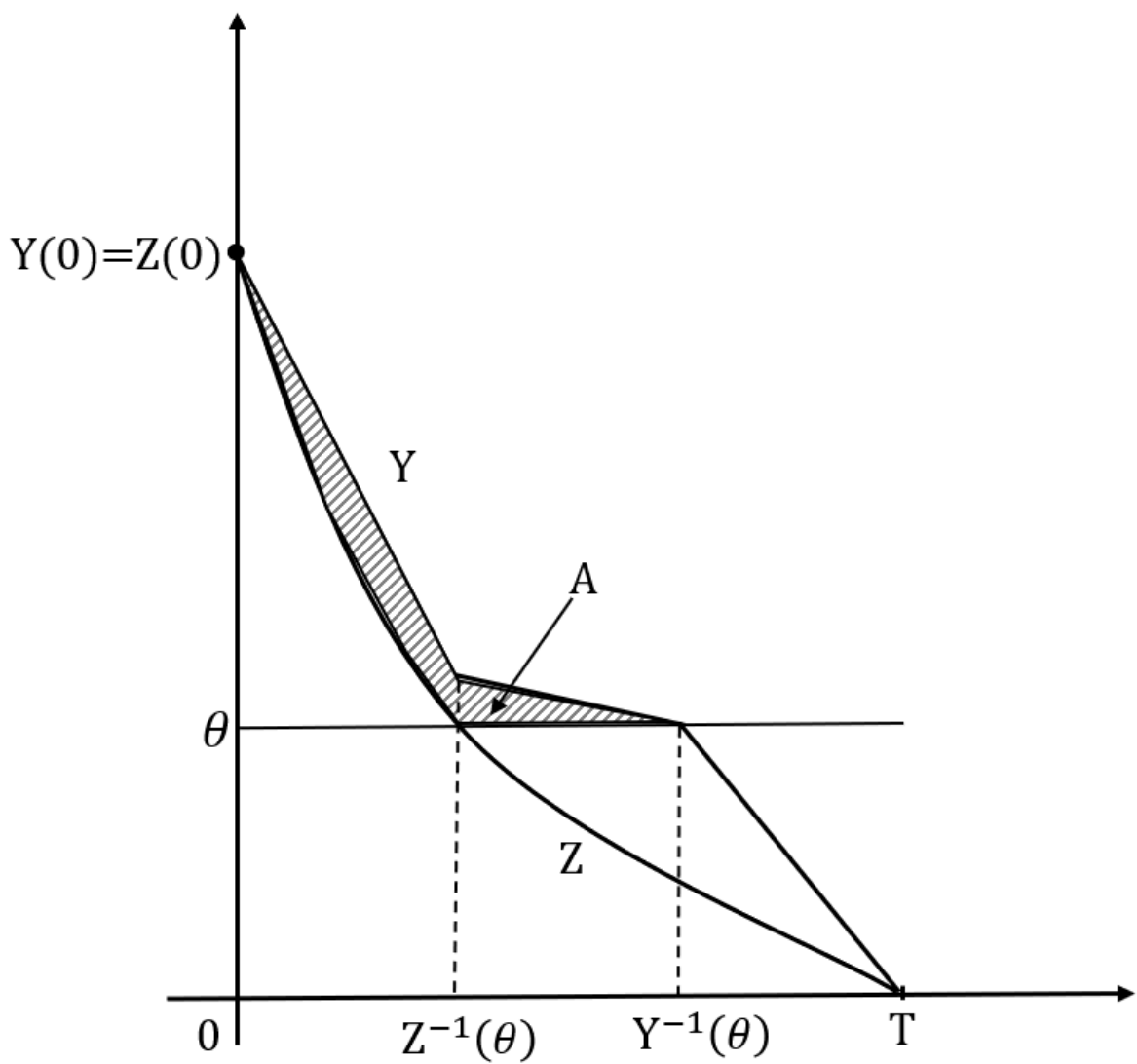

Fig. 4. Illustration for alternative 1

Alternative 2.

We define for each Z ∈ **U**_s and appropriate values of θ

$$\eta_\theta(Z) = \int_0^\theta (Z(s) - Z(\theta))ds$$

The alternative η too is not increasing. We provide an example. Let Y(x) = T-x and let Z(x) be as shown in Fig. 5. Z(x) is linear between (0,T) and (T/2,T/4) and again linear between (T/2,T/4) and (T,0). We have Z ≤ Y on [0,T] and even Z < Y on ]0,T[, but $\eta_\theta(Z) > \eta_\theta(Y)$ for θ = T/2. Indeed:

$$\int_0^{T/2}\left(Z(s) - \frac{T}{4}\right)ds = \int_0^{T/2} Y(s)ds - A - \frac{T^2}{8} = \int_0^{\frac{T}{2}} Y(s)ds - \frac{3T^2}{16}$$

$$> \int_0^{\frac{T}{2}}\left(Y(s) - \frac{T}{2}\right)ds = \int_0^{\frac{T}{2}} Y(s)ds - \frac{T^2}{4}$$

where the area A is shown on Fig. 5.

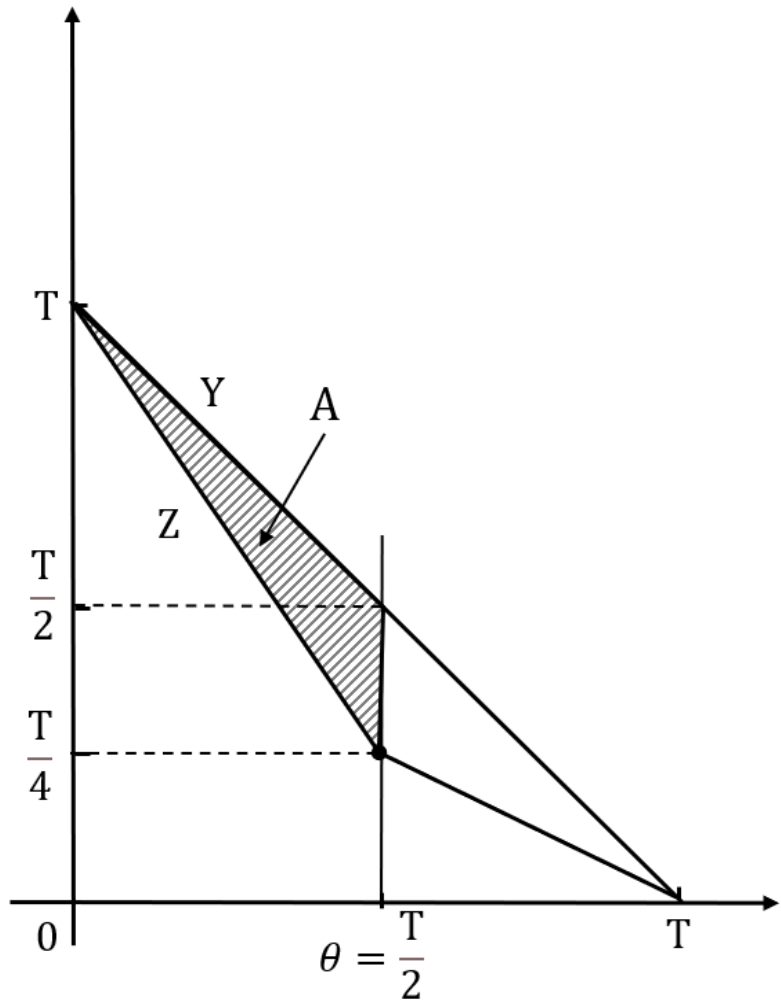

Fig.5 Illustration of alternative 2

## 6. Conclusion

The generalized e-bundle is a non-trivial example of an impact bundle. From a mathematical point of view, we consider it at least as interesting as the generalized h- and g-bundles, because of its use of the inverse function $Z^{-1}$. This is a novel aspect in the study of impact bundles. We note, moreover, that the e-bundle is the first published case that for all admissible values of θ takes the production of the most productive sources into account.

Acknowledgment. The authors thank Li Li (National Science Library, CAS) for drawing the figures in this article. They also thank anonymous reviewers for their comments, leading to some clarifications in the submitted text.

19Egghe, L., & Rousseau, R. (2023). Global impact measures. *Scientometrics,* https://doi.org/10.1007/s11192-022-04553-w

Hairer, E. & Wanner, G. (2008). *Analysis by its History*. Springer (New York). ISBN 978-0-387-77031-4

Hirsch, J.E. (2005). An index to quantify an individual's scientific research output. *Proceedings of the National Academy of Sciences of the United States of America,* 102(46), 16569-16572.

Jin, BH., Liang, LM., Rousseau, R., & Egghe, L. (2007). The R- and AR- indices: Complementing the h-index. *Chinese Science Bulletin*, 52(6), 855-863.

van Eck, N.J., & Waltman, L. (2008). Generalizing the h- and g- indices. *Journal of Informetrics*, 2(4), 263-271. DOI: 10.1016/j.joi.2008.09.004

Yuan, XY., Hua, WN., Rousseau, R., & Ye, F.Y. (2014). A preliminary study of the relationship between the h-index and excess citations / Étude préliminaire de la relation entre l'indice de Hirsch (indice-h) et les citations excédentaires. *Canadian Journal of Information and Library Science / La Revue canadienne des sciences de l'information et de bibliothéconomie* , 38(2), 127-144.

Zhang, C.T. (2009). The e-index, complementing the h-index for excess citations. *PLoS One*, 4(5), e5429. doi:10.1371/journal.pone.0005429

Zhang, C.T. (2010). Relationship of the h-Index, g-Index, and e-Index. *Journal of the American Society for Information Science and Technology,* 61(3), 625–628.